\documentclass[10pt]{article}
\usepackage{tabularx,bm}

\usepackage{amscd}
\usepackage{amsfonts}

\usepackage{yfonts}

\newcommand {\bx} {{\bf x}}

\newcommand  {\bra} {\langle}
\newcommand  {\ket }{\rangle}

\fontsize{12}{1pt} 
  
  \title{Adiabatic definitions of scattering matrix and inclusive scattering matrix}

\date{}	

\begin{document}	
 \author {  A. Schwarz\\ Department of Mathematics\\ 
University of 
California \\ Davis, CA 95616, USA,\\ schwarz @math.ucdavis.edu}

\maketitle
\begin{abstract}
The main goal of present paper is to analyze the adiabatic definition of scattering matrix in the formalism of L-functionals.
 This definition leads to the notion of inclusive scattering matrix closely related to inclusive cross sections.
 We discuss this notion and the relation of our techniques to adiabatic quantum computing.
\end {abstract}

\section{Introduction}
In the geometric approach to quantum theory suggested in \cite {GA} \cite {GA1} the starting point is the set of states. In concrete situations this idea was exploited much earlier in \cite {SCH} where the states of theories obtained by quantization of classical theories were represented as non-linear functionals 
called L-functionals. These functionals are important in the case of an infinite number of degrees of freedom because in this case there exist non-equivalent irreducible representations of CCR and CAR; L-functionals correspond to vectors in all of these representations. (See \cite {MO} and \cite {SP} for the review of the theory of L-functionals. The definitions in the present paper mostly follow \cite {SP}, some of our statements are formulated in \cite {SP} without proofs.)  The diagram techniques in the formalism of L-functionals coincide with the techniques in  Keldysh formalism of non-equilibrium statistical physics ( see for example \cite {KEL}). Later the equations of motion for L-functionals where reinvented with different motivation under the name of 
thermo-field dynamics (TFD) \cite {CU}.

One of the goals of the present paper is to describe (following \cite {T}) the construction of a renormalized scattering matrix in the formalism of L-functionals and to sketch a proof of the relation of this scattering matrix to the conventional scattering matrix and to inclusive cross-sections. The construction presented in \cite {T} is similar to the construction of the conventional renormalized scattering matrix in terms of the adiabatic scattering matrix 
given in \cite {LTS} and the proofs in \cite {T} are based on the ideas of \cite {LTS}. We sketch 
a much simpler proof of the main result of \cite {LTS}; similarly, we  simplify the proofs in \cite {T}. 

The scattering matrix in the formalism of L-functionals coincides with inclusive scattering matrix studied in my recent papers. It can be expressed in terms of amputated GGreen functions on shell \cite {NU}, \cite {PR}, \cite {SP}. This expression follows also from  the results of present paper.

Another goal of present paper is to study the limit of the scattering mat qrix in the formalism of L-functionals(=inclusive scattering matrix) in the limit $\hbar \to 0$ (see also \cite {SING}). Notice that in this paper we are working in the framework of perturbation theory;
the classical limit of scattering of quantum particles corresponding to solitons was considered in \cite {FTS}.

Our constructions are based on the techniques that allow us to obtain stationary states of complicated Hamiltonians from stationary states of simple Hamiltonians using adiabatic (slowly varying)  families of Hamiltonians (adiabatic dressing). These techniques are used also in adiabatic  quantum computing; we consider some generalizations of adiabatic quantum computing.

We start with an explanation of basic definitions of the formalism of L-functionals (Section 2)
in the case of CCR. (The case of CAR is very similar; all results below can be formulated and proved in this case.) To every vector $\Phi$ and, more generally, to every density matrix $K$ 
in a representation space of CCR one can assign a functional
$$L_K(\alpha^*,\alpha)=\mathrm{Tr} e^{-\alpha a^+}e^{\alpha^*a}K$$
where $a_k,a^+_k$ obey CCR, $\alpha a^+=\sum \alpha_ka^+_k, \alpha^*a=\sum \alpha^*_ka_k.$ (This functional can be considered  as positive functional on Weyl algebra.)
We formulate quantum theory in terms of functionals of this kind (L-functionals). In particular, we define evolution operators and adiabatic S-matrix and describe perturbation theory for these objects. In what follows we work in the framework of perturbation theory assuming that 
infrared and ultraviolet divergences are absent. 

In Section 3 we consider adiabatic approximation in conventional quantum mechanics and in the formalism of L-functionals. We discuss  adiabatic quantum computing and its generalizations.

In Section 4 we  remind standard definitions of Green functions and adiabatic Green functions and define corresponding objects in the formalism of L-functionals. We use the term GGreen functions for the analog of Green functions in the formalism of L-functionals; these functions appear also in Keldysh formalism.

In Section 5 we show how one can get the renormalized scattering matrix from the adiabatic S-matrix $\hat S_{a,\Omega}$.(We start with formal Hamiltonian expressed in terms of creation and annihilation operators. Usually, such a Hamiltonian specifies a well-defined operator only after volume cutoff. We define the adiabatic S-matrix $\hat  S_{a,\Omega}$ in finite volume $\Omega$; in  the adiabatic limit the parameter $a$ tends to zero.) In the standard approach taking the adiabatic limit, we obtain a non-renormalized scattering matrix, but we show that by including some phase factors we can obtain 
renormalized S-matrix in the limit. This is very natural - the same factors appear by adiabatic dressing of bare vacuum and bare particles. We assume the absence of infrared and ultraviolet divergences, hence renormalization does not include the renormalization of coupling constants. Our considerations cannot be applied to quantum electrodynamics, however  one can define inclusive scattering matrix in quantum electrodynamics as a limit of inclusive scattering matrices in theories without infrared and ultraviolet divergences. ( The conventional scattering matrix in quantum electrodynamics is trivial because every collision generates an infinite number of soft photons.)

A slightly different approach comes from the 
remark that there is some freedom in the definition of a renormalized scattering matrix. (Notice that this freedom disappears in Lorentz-invariant theories.) Namely,  we say that operators $\hat S$ and $\hat S'$ in Fock space are  $N$-equivalent if $\hat S'=\hat U_1\hat S \hat U_2$ where $\hat U_i$ are unitary operators obeying $\hat U_i a(k)\hat U_i^{-1}=r_i(k) a(k)$.(Here $a(k)$ stands for annihilation operator.)   We say that $N$-equivalence is $S$-equivalence  if $\hat U_2=\hat U_1^{-1}.$ The renormalized scattering matrix $S$ is defined up to $N$-equivalence (up to $S$-equivalence if we require that vacuum vector and one-particle states are invariant with respect to $\hat S$). The results of Section 5 lead to the following definition of scattering matrix:

An operator $\hat S$ in Fock space is a renormalized scattering matrix if it can be represented as a limit 
$$\hat S=\lim_{a\to 0}\lim_{\Omega\to \infty}\hat S'_{a,\Omega}$$
where the operator  $\hat S'_{a,\Omega}$ is $N$-equivalent to adiabatic $S$-matrix $\hat S_{a,\Omega}.$

In Section 6 we prove analogs of the results of Section 5 in the formalism of L-functionals.
The proofs are similar but simpler because the volume cutoff is unnecessary. The scattering 
matrix in the formalism of L-functionals denoted by $S$ is related to the conventional scattering matrix $\hat S$ by the formula 
$$SL_K=L_{\hat SK\hat S^*}.$$
It is easy to express inclusive cross-sections in terms of matrix elements of $S$ using this formula. This fact justifies the name ''inclusive scattering matrix" for $S.$

We show that inclusive scattering matrix can be expressed in terms of amputated GGreen functions on shell.
We conjecture that this expression works also in the case of quantum electrodynamics (the infrared divergences do not appear in the formalism of L-functionals).

Notice that  there exist other definitions of inclusive scattering matrix  ( see \cite {SP} and references therein).
These definitions lead to the same expression of inclusive scattering matrix in terms of  amputated GGreen functions on shell.

Finally, in Section 7 we analyze the inclusive scattering matrix in the limit $\hbar \to 0.$

\section {L-functionals}

In the algebraic approach to the quantum theory, we start with  $*$-algebra (unital associative algebra with involution).
States are interpreted as positive linear functionals on such an algebra. (One says that a linear functional $\omega$ is positive if $\omega (A^*A\geq 0)$. The functional $\omega$ is normalized if $\omega(1)=1$.)

Among the most important $*$-algebras are Weyl algebra and Clifford algebra (the representations of these algebras can be interpreted as representations of canonical commutation and anticommutation relations). We define L-functionals as positive
linear functionals on one of these algebras.

Let us write  canonical commutative relations (CCR) in  the form: 
\begin{equation}\label {CCC}
[a(k), a^+(k')]=\hbar\delta (k,k'),
[a(k), a(k')]= [a^+(k), a^+(k')]=0,
\end {equation}
where $k,k' \in M$. We assume that $M$ is a measure space. Recall that in these relations we are dealing with generalized functions; in other words, we should work with formal expressions  $fa=\int dk f(k)a(k)$, $ga^+=
\int dk g(k)a^+(k)$ where $f$ and $g$ belong to some class of functions (test functions).  In the discrete case, the integral is understood as a  sum, and the  $\delta$- function as the Kronecker symbol:
\begin{equation}\label {CCRD}[a_k,a^+_{k'}]=\hbar\delta_{k,k'},
[a_k, a_{k'}]=[a^+_k,a^+_{k'}]=0.
\end{equation}

We say that $a^+(k)$  are creation operators and $a(k)$ are annihilation operators.

The canonical anticommutation relations (CAR) can be written in the form
\begin{equation}\label {CAR}[a(k), a^+(k')]_{+}=\hbar\delta (k,k'), [a(k), a(k')]_+= [a(k)^+, a(k')^+]_+=0, \end{equation}
In what follows we work with CCR, but the case of CAR is very similar.

Together with notations $fa, ga^+$ we will use the notations $a(f)=fa, a^+(g)=ga^+$. In these notations, CCR have the form
\begin {equation}
\label {CCR}
[a(f), a(g)]=[a^+(f),a^+(g)]=0, [a(f), a^+(\bar g)=\hbar\bra f,g \ket 
\end{equation}
CCR written in the form (\ref{CCR}) make sense if $f, g\in \cal E$ where $\cal E$ is a complex pre Hilbert space with inner product $\bra f,g \ket .$  The elements of $\cal E$ play the role of test functions.

Let us consider a representation of CCR. We assume that the operator $i(f^*a-fa^+)$ is self-adjoint (this is a rigorous expression of the assumption that
 the generalized operator functions $a(k)$ and $a^+(k)$ are Hermitian conjugate).

 We  define 
   the L-functional corresponding to a density matrix $K$ in the representation of CCR by the formula:
\begin{equation}
\label {LLL}
L_K(\alpha^*,\alpha)=\mathrm{Tr} e^{-\alpha a^+}e^{\alpha^*a}K.
\end{equation}
Here $\alpha(k)$ is a function on $M.$

We can write this definition in the form
\begin{equation}
\label {LL}
L_K(\alpha^*,\alpha)=\mathrm{Tr} W_{\alpha}K.
\end{equation}
where
 \begin {equation}\label {WWW}
 W_{\alpha}=e^{-\alpha a^+}e^{\alpha^*a}
 \end{equation}

An easy formal calculation shows that
 \begin{equation} \label {EXP}
W_{\alpha}W_{\beta}= e^{-(\alpha^*,\beta)} W_{\alpha+\beta} ,
\end{equation} 
 
One can say that L-functional is just a generating functional for correlation functions. 

We assume that $\alpha$ is a square-integrable function, then the expression (\ref {WWW}) is well-defined because it coincides with a unitary operator
$e^{-\alpha a^+ +\alpha^*a}$ up to a finite numerical factor. However, the case when $\alpha$ is not square integrable is also interesting; it is important for applications to string theory.

One can define Weyl algebra $\cal W$ as a $*$-algebra generated by $a(k), a^+(k)$  (or, more precisely by $fa=\int dk f(k)a(k)$, $ga^+=
\int dk g(k)a^+(k)$) obeying CCR.
We  modify slightly the definition of $\cal W$  assuming that it is generated by
  $a(k), a^*(k),$ and $W_{\alpha}$ obeying  (\ref {CCC}), (\ref{EXP})
  and relations
  \begin{equation} \label{CR}
  a(k)W_{\alpha}=W_{\alpha}a(k)-\alpha(k),a^+(k)W_{\alpha}=W_{\alpha} a^+(k)-\alpha^*(k)
  \end{equation}
 We can consider $\cal W$ as a version of Weyl algebra.
  To specify a topology in $\cal W$ one  can consider a representation of CCR containing a cyclic vector $\theta$ that obeys $a(k)\theta)=0$
  (Fock representation) and use, for example, strong topology in the space of operators.  Then linear combinations of elements $W_{\alpha}$ are dense in $\cal W$.  (The operators $fa,ga^+$ are defined on a dense subset of Fock space.  Strong convergence of a sequence of operators $A_n$ to the operator $A$ is defined as convergence of $A_nx$ to $Ax$ for every element $x$ of this subset. The choice of topology in $\cal W$ is irrelevant for the considerations below.)

Let us denote by $\cal L$ the space of continuous linear functionals on $\cal W.$  Such a functional is completely determined by its values on generators $W_{\alpha}$; we denote these values by 
${\bf L}(\alpha^*,\alpha).$  In other words, a linear functional $L$  can be represented by a non-linear functional  ${\bf L}(\alpha^*,\alpha)=L(W_{\alpha}).$

The action of the Weyl algebra $\cal W$ on the space $\cal L$ is realized by the operators $b$ and $b^+$  whose action on functionals $L_K$ corresponds to the multiplication of the density matrix by the operators $a^+$ and $a$ from the right:
$$b(k)L_K = L_{Ka^+(k)}, \;\;\; b^+(k)L_K = L_{Ka(k)}.$$
It is easy to check that these operators satisfy the canonical commutative relations and can be represented in the following form
\begin{equation} \label {B}
b^+(k)=-\hbar c_2^+(k)+c_1(k), \;\;\; b(k)=-c_2(k), 
\end{equation}
where $ c^+_i (k)$ are multiplication operators by $\alpha(k)^*$ for $i=1$ and by $\alpha(k)$ for $i=2$, and $c_i(k)$ are derivatives  with respect to $\alpha^*(k)$ and $\alpha (k)$. 

An alternative action of $\cal W$ on $\cal L$ is realized by operators, whose action on functionals $L_K$ corresponds to the multiplication of the density matrix by operators $a$ and $a^+$ from  the left:
$$\tilde b(k)L_K = L_{a(k)K},\;\;\;\tilde b^+(k)L_K = L_{a^+(k)K} $$
These operators also obey the canonical commutative relations. They can be represented in the form:\begin{equation}\label{BB}
\tilde b^+(k)=\hbar c^+_1(k)-c_2(k),\;\;\;\tilde b(k)=c_1(k).
\end{equation} 

It follows from  (\ref {B}) and (\ref{BB}) that

\begin{equation}\label{BBB} \tilde b^+(k)-b(k)=\hbar  c^+_1(k),\tilde b(k)-b^+(k)=\hbar c_2^+(k)
\end{equation}

 Notice that  $c_i(k), c^+_i (k)$ satisfy CCR, but with $\hbar=1$, therefore taking the limit $\hbar\to 0$ it is convenient to  use formulas (\ref {B}), (\ref{BB}), (\ref {BBB}).

Thus, there are two commuting actions of Weyl algebra on $\cal L$.  (In the terminology of physicists we have a doubling of fields.)

The above definitions are particular cases of the following general construction that can be applied to any  $*$-algebra $\cal A$. For every  element $B\in \cal A$ we can define two operators acting in the space of linear functionals on $\cal A$. One of them transforms a functional $\omega (A)$ into the functional 
  $\omega (AB)$; it will be denoted by the same symbol $B$, another transforms  this functional into the functional  $\omega (B^*A)$; it will be denoted by $\tilde B.$
  
  In these notations two operators corresponding to $a(k)$ are $b^+(k),  \tilde b^+(k)$, two operators corresponding to $a^+(k)$ are $b(k), \tilde b(k).$ If $\phi$ is a self-adjoint linear combination of $a(k)$ and $a^+(k)$ then it follows from (\ref {BBB}) is proportional to $\hbar.$
  
  Notice that 
  \begin{equation}\label {TIL}
  (\tilde B B\omega)(A)=\omega (B^*AB)
\end{equation}
hence the operator $\tilde B B$ transforms a positive functional into positive functional.

Consider now a formal Hamiltonian $\hat H$: 
 
\begin{equation}
\label{hh}
\hat H=\sum _{m,n}\int \Gamma_{m,n}(k_1, ...k_m|l_1,...,l_n)a^+(k_1)...a^+(k_m)a(l_1)...a(l_n)d^mkd^nl
\end{equation}
expressed in terms of creation and annihilation operators and presented in the normal form (i.e., all creation operators are moved to the left). In many interesting cases, the formal expression (\ref{hh}) does not define a self-adjoint  operator in Fock space. 
In the algebraic approach, formal Hamiltonians may not make sense as operators, but the corresponding equations of motion may make sense.

The Hamiltonian $\hat H$ induces two formal operators acting on $\cal L$:
\begin{equation}
\nonumber
\hat H=\sum _{m,n}\int \Gamma_{m,n}(k_1, ...k_m|l_1,...l_n)b^+(k_1)...b^+(k_m)b(l_1)...b (l_n) d^mkd^nl
\end{equation}
\begin{equation}
\nonumber
\tilde H=\sum _{m,n}\int \Gamma_{m,n}(k_1, ...k_m|l_1,...,l_n)\tilde b^+(k_1)...\tilde b^+(k_m)\tilde b(l_1)...\tilde b (l_n) d^mkd^nl
\end{equation}
One of them is denoted by the same symbol $\hat H$, the other is denoted by $\tilde H$. 

Now we can write the equation of motion for the L-functional  $L(\alpha^*,\alpha)$:
\begin{equation}
\label{HH}
\frac{dL}{dt}= {H} L=\frac {1}{i\hbar}(\tilde HL - \hat HL),
\end{equation}
where we introduced  the notation $ H=\frac {1}{i\hbar}(\tilde H - \hat H).$ We say that $H$ is a ''Hamiltonian".

If we consider a translation- invariant Hamiltonian (\ref{hh}), then in the momentum representation the coefficients  $\Gamma_{m,n}$ contain $\delta$-functions $\delta (k_1+...+k_m-l_1-...-l_n)$ (they express the momentum conservation law).

Using (\ref{B}), (\ref {BB}) we can express the right-hand side of  (\ref{HH}) in terms of  $c_i(k), c^+_i (k)$. It is easy to check that in RHS of (\ref{HH}) the terms that do not contain $\hbar$ (=the terms that do not contain operators $c_i^+(k$)) cancel. This means that {\it the equation (\ref{HH}) has a limit as $\hbar\to 0.$}

Let us assume now that the formal Hamiltonian $\hat H$ is represented as  a sum of quadratic Hamiltonian $\hat H(0)$ and interaction Hamiltonian $\hat H_{int}=g\hat V.$ We  take $\hat H(0)$ in the form
\begin{equation} \label {H0} 
\hat H(0)=\int \epsilon (k) a^+(k)a(k)dk.
\end{equation} We always assume that every term in the interaction Hamiltonian contains at least three creation and annihilation operators (i.e. the coefficient functions of the interaction Hamiltonian vanish if $m+n\leq 2$). 

If we consider translation-invariant Hamiltonian we  assume that $k=({\bf k}, s)$ where $\bf k$ stands for momentum variable and $s$ stands for discrete index (then the integral over $k$ should include summation over discrete index). However, for simplicity, we suppose that the discrete index is absent.

To avoid infrared divergences we assume that $\epsilon (k+l)<\epsilon(k)+\epsilon(l)$.
This condition means that energy and momentum conservation laws guarantee the stability of particles. If the discrete index is present we should suppose that the stability of particles is guaranteed by some conservation laws (by energy, momentum, and some other conservation laws).

If the formal Hamiltonian (\ref{hh})  specifies a self-adjoint operator in the Fock representation of CCR then we can define in a standard way the evolution operator $\hat U(t,t_0)$, the evolution operator in the interaction picture $\hat S (t, t_0)$ and the adiabatic $S$-matrix  $\hat S_{a}$.  

 Namely,  we denote by $\hat U(t, t_0)$ the operator transforming a vector at the moment $t_0$ into the vector at the moment $t$ ( the evolution operator). To define the adiabatic $S$-matrix we introduce
the evolution operator  $\hat U_{a}(t, t_0)$ by  the equation
$$i\hbar \frac {d} {dt }\hat U_{a}(t, t_0)= (\hat H(0)+h(a t)\hat H_{int})\hat U_{a}(t, t_0)$$
with initial data $\hat U_{a}(t_0, t_0)=1.$ (We assume that $h(t)$ is a smooth function obeying $h(0)=1, h(t)=h(-t)$ and fast decreasing as $t\to \infty$. The Hamiltonian is changing slowly in the limit $a\to 0.$) 

The operator $\hat S_{a}(t,t_0)$ is defined by the formula $$\hat S_{a}(t,t_0)= e^{\frac {i}{\hbar}\hat H(0)t}\hat U_{a}(t, t_0)e^{- \frac {i}{\hbar}t \hat H(0)t_0}.$$
 It satisfies the equation
 $$i\hbar \frac {d} { dt} \hat S_{a}(t, t_0)= h(at)\hat H_{int}(t) \hat S_{a}(t, t_0)$$ with initial data $ \hat S_{a}(t_0, t_0)=1.$ (We use the notation $A(t)= e^{\frac {i}{\hbar}\hat H(0) t}A e^{-\frac {i}{\hbar}\hat H(0) t}.$)
  The adiabatic $S$-matrix $\hat S_{a}$ is defined as $\hat S_{a}(\infty, -\infty).$

Using  the formula
\begin{equation}\label{SS}
\hat S_{a}(t,t_0)=T\exp(-\frac {i}{\hbar}\int _{t_0}^td\tau h(a \tau) \hat H_{in\tau}(\tau))
\end{equation}
one can construct the perturbation series  for  $\hat S_{a}(t,t_0)$ and for adiabatic S-matrix.

If the formal Hamiltonian (\ref{hh}) is translation-invariant then usually it does not specify an operator in a representation of CCR, however, it specifies an operator  $\hat H_{\Omega}$ in the Fock representation after volume cutoff (i.e. after replacement of the continuous parameter  $k$  by a discrete parameter that runs over a lattice).  We use a subscript $\Omega$ to emphasize that we are working in finite volume. ( Without volume cutoff the Hamiltonian 
(\ref {hh}) can define an operator in Fock space  only if it does not have terms containing only creation operators.)

The corresponding ''Hamiltonian" (\ref{HH}) in the space of L-functionals leads to well-defined equations of motion
(at least in the framework of perturbation theory) even without volume cutoff if the perturbation theory for the Hamiltonian $(\ref {hh})$ is well-defined after volume cutoff. We can define  corresponding evolution operators 
$U(t,t_0)$ and  $U_a(t,t_0)$, the evolution operator in the interaction picture  and the analog of adiabatic S-matrix  $\hat S_a$ denoted by $S_a.$

In particular, if the ''Hamiltonian" $H_a=H(0)+h(a t)H_{int}$  governs the evolution of $L$-functional  then in the interaction picture we have
\begin{equation}
\label {SSS}
 S_{a}(t,t_0)=T\exp(\int _{t_0}^td\tau h(a \tau) H_{int}(\tau)).
 \end{equation}
The perturbation theory for operators $S_{a}(t,t_0)$ and $S_a=S_a(+\infty,-\infty)$  can be derived from this formula in the standard way.

{\it If the operators $\hat S_{a}(t,t_0)$ are well-defined they are related  to the operators $S_{a}(t,t_0)$ by the formula}
\begin{equation}\label{HAT}
S_{a}(t,t_0)L_K=L_{ \hat S_{a}(t,t_0)K\hat S_{a}(t,t_0)^*}
\end{equation}
Notice, however, that in many cases operators $ S_{a}(t,t_0)$ are well-defined, but the operators $\hat S_{a}(t,t_0)$ do not make sense.

{\it In what follows we assume that $\hbar=1$ if we do not consider the semiclassical limit.}

\section { Adiabatic approximation and adiabatic quantum computing}

Let us start with some well-known facts about adiabatic approximation in quantum mechanics.
We consider a family of Hamiltonians $\hat H(g)$ acting in Hilbert space $\cal H$ and the Schr\o  dinger equation
\begin {equation} \label {SCH} i\frac {d\psi}{dt}=\hat H(g(t))\psi \end{equation}
where $g(t)$ is a slowly varying function (this means that we can neglect the derivative of $g(t)$ with respect to time $t$).

Let us suppose that  $\phi(g)$ is an eigenvector of $\hat H(g)$:

$$\hat H(g)\phi(g)=E(g)\phi (g).$$
(We assume that $\hat H(g)$ and $\phi(g)$ are smooth functions of $g$ and $\phi(g)$ is a normalized vector. The function $\phi(g)$ is defined only up to multiplication by a numerical smooth function $C(g)$; we get rid of this ambiguity by fixing $\phi(0)$ and imposing the condition
$\bra\frac {d\phi}{dg}, \phi(g)\ket=0.$) 

Then  in adiabatic approximation $e^{-i\beta (g(t))}\phi(g(t))$ where $\frac {d\beta}{dg}=E(g)$
obeys (\ref {SCH}).  (Recall that in adiabatic approximation we neglect $\dot g(t).$) If $E(g)$ is an isolated simple eigenvalue
and $\hat H(g)=\hat H(0)+\hat H_{int}=\hat H(0)+g\hat V$ we can say that
\begin {equation}\label {PHI}
\phi(1)=\lim_{a\to 0} \exp(i\int_{-\infty}^0 d\tau(E(h(\tau)-E(0))) \hat S_a(0,-\infty)\phi(0)
\end {equation}
We will say that the state $\phi(1)$ is obtained from the state $\phi(0)$ by adiabatic dressing.

More generally
\begin{equation} \label {PH}
\phi(g)=\lim_{a\to 0}\exp (i\int_{-\infty}^0 d\tau(E(h(\tau)-E(0)))\exp(\int_0^td\tau E(h(\tau)))\exp(-i\hat H(0)t)\hat S_a(t,-\infty)\phi(0)
\end{equation}
In this formula, we assume that $h(t)=g.$

In particular, the formula (\ref {PHI})  allows us to express the ground state of the Hamiltonian $\hat H(1)=\hat H(0)+\hat V $ in terms of the ground state $\phi=\phi(0)$  of the Hamiltonian $\hat H(0).$ 

Notice we assume that  $\phi (g)$ is a simple isolated  eigenvalue of the Hamiltonian $\hat H(g)$. In particular, if $\phi(g)$ is a ground state the Hamiltonian $\hat H(g)$ should have a gap (i.e.the energy of the ground state should be separated from other energy levels by a positive number).
 
The situation with adiabatic approximation in L-functionals' formalism is similar but simpler. Let us denote by $L(g)$   a stationary state of the ''Hamiltonian" $H(g)$; we assume that both the L-functional $L(g)$ and the ''Hamiltonian" $H(g)$ are smooth functions of $g$. Then $L(g(t)$ is an approximate solution of the equation
\begin {equation} \label {LT}
\frac {dL}{dt}=H(g(t))L
\end{equation}
if we can neglect $\dot g(t).$ ( We say that $L(g(t))$ solves (\ref{LT}) in adiabatic approximation.)

If  $L(0)$ is a ground state of the ''Hamiltonian" $H(0)$ then the ground state of the ''Hamiltonian"
$H(0)+V$  can be represented in the form 
$$ \lim_{a\to 0} U_a(0,-\infty) L(0)$$
where $U_a$ denotes the evolution operator for the time-dependent ''Hamiltonian" $H(0)+h(at)V$
( The ground state of ''Hamiltonian" is the L-functional describing the ground state of the corresponding Hamiltonian, therefore this formula is equivalent to the formula (\ref {PHI}).)  Instead of $U_a$ one can write $S_a$ (the evolution operator in the interaction picture) in this formula.

More generally,
if $\omega_0$ is a stationary state of the ''Hamiltonian" $H(0)$ and there exists a limit
\begin{equation} \label{LIM}
 \lim_{a\to 0} U_a(0,-\infty) \omega_0
\end {equation}
then this limit is a stationary state of the ''Hamiltonian" $H(0)+V.$ Again the operator $ U_a(0,-\infty)$ can  constructed be replaced by $ S_a(0,-\infty)$ in this formula.

We say that the state (\ref{LIM}) is obtained from $\omega_0$ by adiabatic dressing.

Notice that the limit (\ref {LIM}) not necessarily exists. However, we should expect that it exists in the case when $\omega_0$ is an equilibrium state. ( When the Hamiltonian changes adiabatically an equilibrium state remains an equilibrium state, maybe with different temperature).

The formula (\ref {PHI}) lies in the basis of adiabatic quantum computing \cite {A},\cite{B}. To use it one should reduce the problem we are interested in to a problem of finding of ground state of some Hamiltonian $\bar H$. In the paper \cite{D} such a reduction was found in the case when our problem is solved by means of gate-based quantum computer; moreover it was shown in this paper that the Hamiltonian $\bar H$ can be analyzed by means of 
adiabatic methods. The authors of \cite {D} conclude that  {\it 
The model of adiabatic computations is polynomially equivalent to the model of standard (gate-based) quantum computations.}

Using (\ref {LIM}) we can apply adiabatic quantum computing in the case when our problem can be reduced to finding an equilibrium state of some Hamiltonian.

\section{Green functions and generalized Green functions}
The formula (\ref {PHI}) 
 permits us to express physical Green functions
\begin {equation}\label {G}
\bra T(\hat{\bf  A}_1(t_1)...\hat {\bf A}_n(t_n)) \Phi, \Phi\ket
\end{equation}
where $T$ stands for the chronological product of Heisenberg operators $\hat {\bf A}(t)= e^{i\hat Ht}Ae^{-i\hat H t}$(times decreasing) and $\Phi=\phi(1)$ denotes the ground state of the Hamiltonian $H(1)$
as limits of adiabatic Green functions
\begin{equation}\label {GA}
\frac{\bra \phi|T(\hat A_1(t_1)...\hat A_n(t_n)\hat S_a)|\phi\ket}{\bra\phi|\hat S_a|\phi\ket}=
\frac{\bra T(\hat A_1(t_1)...\hat A_n(t_n) e^{-i\int_{-\infty}^{\infty} d\tau h(a\tau)\hat H_{int}(\tau)})\phi,\phi\ket}
{\bra \hat S_a(\infty,-\infty)\phi,\phi\ket}
\end {equation}
when $a\to 0.$ 

Especially important are the two-point Green functions $\bra T(\hat{\bf  A}_1(t_1)\hat {\bf A}_2(t_2)) \Phi, \Phi\ket.$

For translation-invariant Hamiltonian, we can define  Green functions  by the formula
\begin {equation}\label {GGG}
\bra T(\hat{\bf  A}_1(x_1,t_1)...\hat {\bf A}_n(x_n,t_n)) \Phi, \Phi\ket
\end{equation}
where $\hat {\bf A}(x,t)=e^{iPx}Ae^{-iPx}$ and $P$ stands for momentum operator (generator of translations).   Functions (\ref{GGG}) are Green functions in $(x,t)$-representations; taking Fourier transforms we obtain Green functions in $(p,t)$- and $(p,\omega)$-representations. LSZ formula allows us to express the scattering matrix in terms of Green functions (see, for example, \cite {SP}).

If $\cal H$ is a representation of canonical commutation relations  (\ref{CCC}) and the operators $A_i$ are linear with respect to $a,a^+$ this formula allows us to construct diagram techniques for the calculation of Green functions. We start with the numerator of the expression for the adiabatic Green functions; for the numerator, the vertices are coefficients of $\hat H_{int}$ multiplied by $h(at)$, and propagators are
two-point Green functions for $\hat H(0)$. The diagrams for the denominator are the same,  but they do not have external vertices. To take into account the denominator we exclude diagrams with subdiagrams without
external vertices.  To get diagrams for physical Green functions we take the limit $a\to 0$; this means that we should exclude the factor $h(a\tau)$ from vertices.

Let us consider physical Green functions in the formalism of L-functionals (GGreen functions). If $A_i$ are linear operators in the space $\cal L$ of linear functionals on Weyl algebra (or more generally on any $*$-algebra $\cal A$) we define generalized Green functions (GGreen functions) in the state $\omega$ as the value  of the functional
$$T({\bf A}_1(t_1)...{\bf A}_n(t_n))\omega$$
at the unit element of the algebra or, more generally, at an arbitrary translation-invariant element $\beta$ of the algebra. In what follows we assume that $\beta=1$. (Here ${\bf A}(t)$ stands for the Heisenberg operator $e^{Ht}Ae^{-Ht}$ where $H$ is the 
''Hamiltonian" .) As we know, every element of $*$-algebra specifies two operators acting in $\cal L$; we take as $A_i$ some of these operators.  

If we are working with Weyl algebra and represent elements of the space $\cal L$ by functionals $L(\alpha^*,\alpha)$ then taking the value of the linear functional at the unit element is equivalent to calculating $L(\alpha^*,\alpha)$ at the point $\alpha^*=\alpha=0.$

GGreen functions are closely related to Green functions in Keldysh formalism that are defined by the formula $\omega (MN)$ where $M$ is a chronological product of Heisenberg operators and $N$ is an antichronological product of Heisenberg operators (see \cite{SP}).

If $\omega$ is a stationary state of the ''Hamiltonian" $H(0)+V$ that can be represented by the formula (\ref{LIM}) (is obtained from 
we can use (\ref{SSS}) to represent the GGreen function as a limit of  adiabatic GGreen functions
$$(T(A_1(t_1) ...A_n(t_n)  e^{\int_{-\infty}^{\infty} d\tau h(a\tau) H_{int}(\tau)})\omega_0)(1)$$
as $a\to 0$. (Notice that $S_a(\infty, -\infty)\omega_0=\omega_0$ therefore we do not have a denominator in the expression for the adiabatic GGreen function.) If $\cal A$ is a Weyl algebra and  $A_i$ are linear combinations of operators in $\cal L$
corresponding to creation and annihilation operators we obtain diagram techniques for calculations of adiabatic GGreen functions in a standard way. The propagators are two-point GGreen functions for the ''Hamiltonian"$H(0)$, there are two types of vertices. Taking the limit $a\to 0$ we obtain the diagram techniques for GGreen 
functions in the state $\omega$; these techniques coincide with Keldysh diagram techniques (see \cite {SP} for more details).
\section{ Adiabatic definition of scattering matrix}

Let us consider formal translation-invariant Hamiltonian (\ref{hh})  represented in the
form $\hat H=\hat H(0)+\hat V$ where $H(0)$ is defined by the formula (\ref{H0}) where $\epsilon(k)>0$ and
 \begin{equation}\label {OM}
\epsilon(k_1+...k_n)<\epsilon(k_1)+...+\epsilon(k_n).
\end{equation}
Here $k$ stands for the momentum variable. (Our considerations can be generalized to  the case when in addition to the momentum variable $k$ we have a discrete variable, for example, spin; however, we do not consider this case in the present paper.)

Together with this Hamiltonian, we consider a family of Hamiltonians $\hat H(g)=\hat H(0)+\hat H_{int}=\hat H(0)+g\hat V$ depending on parameter $g.$ Our considerations will be based on perturbation theory with respect to $g.$

 We assume that by applying volume cutoff to $\hat H(g)$ we obtain 
a well-defined self-adjoint operator $\hat H_{\Omega}(g)$ in Fock space ${\cal F}_{\Omega}$
(in the Fock representation of CCR (\ref{CCRD}) where $k$ runs over a lattice). (It is sufficient to suppose that
coefficients $ \Gamma_{m,n}(k_1, ...k_m|l_1,...l_n)$ where $m+n>2$ are   fast decreasing as $k_i, l_j\to \infty$.)  

The Hamiltonian $\hat H_{\Omega}(0)$ has a ground state $\theta$ (the Fock vacuum). We assume that the Hamiltonian $ \hat H_{\Omega}(g)$ has a non-degenerate ground state 
$\Theta_{\Omega}(g)$ with energy $E_{\Omega}(g)$  and that there exists a gap between this energy and other energy levels that does not tend to zero as $\Omega\to\infty.$
  
 One-particle states $|k\ket =a_k^+\theta$ of the Hamiltonian $\hat H_{\Omega}(0)$ can be characterized as normalized vectors with minimal energy in a subspace of elements with fixed momentum $k$ (this characterization follows from (\ref {OM}).  One-particle states 
 $\phi_{\Omega}(k,g)$ of the Hamiltonian
 $\hat H_{\Omega}(g)$ can be characterized in the same way; their energy will be denoted by
 $E_{\Omega}(k,g)$ and  the difference $E_{\Omega}(k,g)-E_{\Omega}(g)$ will be denoted $\epsilon_{\Omega}(k|g).$ 
 
 Our definition specifies normalized vectors  $\Theta_{\Omega}(g)$  and $\phi_{\Omega}(k,g)$ only up to numerical factor; to get rid of this ambiguity we can impose the conditions
 $$\bra \Theta_{\Omega}(g), \frac {\partial \Theta_{\Omega}(g)}{\partial g}\ket=0,\bra \phi_{\Omega}(k,g); \frac {\partial \phi_{\Omega}(k,g)}{\partial g}\ket=0. $$ 
 
  Notice that the momentum operator has a discrete spectrum in finite volume.  We use the notation $|p_1,...,p_n\ket$ for the standard basis in Fock space (in finite volume $p_i$ runs over a lattice). This basis consists of eigenvectors of the momentum operator.
 
 Green functions of the formal Hamiltonian (\ref {hh}) can be defined as limits of Green functions
 of Hamiltonians $\hat H_{\Omega}$ as $\Omega \to \infty.$ (The momentum variable runs over a lattice; we should extend the Green function in $(p,t)$-representation to a function of continuous variables and take a limit in the sense of generalized functions.) These Green functions can be interpreted as Green functions of field theory in appropriate Hilbert space (see \cite{MO}); this fact allows us to define the scattering matrix and apply the LSZ formula to  calculate it.
 
 This scattering matrix $\hat S$ can be expressed in terms of the adiabatic scattering matrix in finite volume
 denoted by $\hat  S_{a,\Omega}.$ This expression can be represented in several different ways.
 
 Let us introduce the following notations.
 
 $$ A_{a,\Omega}=\bra\theta|\hat S_{a,\Omega}|\theta\ket^{-\frac 1 2}$$
$$B_{a,\Omega}(p)=\bra p|\hat S_{a\Omega}|p\ket^{\frac 1 2}$$
$$r_{a,\Omega}(p)=i\ln(A_{a,\Omega}B_{a,\Omega}(p)).$$

We prove that {\it he renormalized scattering matrix $\hat S$ can be expressed in terms of the adiabatic scattering matrix in finite volume $\Omega$ in the following way}
\begin{equation}\label {SSS}
\hat S=\lim_{a\to 0}\lim_{ \Omega\to\infty}\frac{ \hat U_{a,\Omega}\hat S_{a,\Omega}\hat U_{a,\Omega}}{\bra\theta|\hat S_{a,\Omega}|\theta\ket}
\end{equation}
{\it where $$\hat U_{a,\Omega}=e^{i\sum_kr_{a,\Omega}(k)a^+(k)a(k)},$$
the limit is understood as the convergence of matrix elements in the sense of generalized functions.}

In terms of matrix elements, the relation (\ref{SSS}) can be represented in the form
$$\bra p_1,...,p_m|\hat S|q_1,..,q_n\ket=$$
$$\lim_{a\to 0}\lim_{ \Omega\to\infty}\frac{\bra p_1,...,p_m|\hat S_{a,\Omega}|q_1,..,q_n\ket A_{a,\Omega}^{2-m-n} }{B_{a,\Omega}(p_1)...B_{a,\Omega}(p_m)B_{a,\Omega}(q_1)...B_{a,\Omega}(q_n)}$$

Notice that  it follows from adiabatic perturbation theory (\ref{PHI})  that for $a\to 0$ 
we have 
$$ A_{a,\Omega}\approx e^{-i\int_{\infty}^0 d\tau E_{\Omega}(h(\tau))},$$
$$B_{a,\Omega} (p)\approx e^{i\int_{-\infty} ^0d\tau (E_{\Omega}(p,h(\tau))-\epsilon(p))},$$

  This means that in (\ref{SSS}) one can take 
\begin{equation}\label{R}
r_{a,\Omega}=\int_{-\infty}^0(\epsilon_{\Omega}(k|h(\tau))-\epsilon(k))d\tau.
\end{equation}
 We prove (\ref{SSS}) with this definition of  $r_{a,\Omega}$; all other statements can be derived from this result.

The proof is based on the representations of adiabatic scattering matrix $\hat S_{a,\Omega}$ and renormalized scattering matrix $\hat S$ in terms of diagrams where vertices are 1PI diagrams and propagators are physical two-point Green functions depicted by thick lines. (Recall that one-particle irreducible or 1PI  diagrams are defined as diagrams that remain connected if we cut one of internal edges.)  In  $(p,t)$-representation diagrams for $\hat S_{a,\Omega}$ contain a factor $h(at_1)...h(t_n)$ that tends to $1$ as $a\to 0$. For 1PI diagrams this implies that diagrams
 for $\hat S_{a,\Omega}$ tend to corresponding diagrams for $\hat S$.  (It is easy to verify this statement going to $(p,\omega)$-representation and noticing that in this representation  1PI diagrams for $\hat S$ have no poles  for generic momenta $p$ \cite{STA}.  We can work also in $(p,t)$-representation; then it follows from \cite{STA} that 1PI diagrams decrease faster than any power of $t$ as $t\to \pm\infty$ for open dense set of momenta $p$.) Similarly, internal propagators for $\hat S_{a,\Omega}$ tend to corresponding internal propagators for $\hat S$. (Internal propagators  for $\hat S$ have no poles; a pole would correspond to a physical particle and to a different scattering process.)
 It remains to analyze external propagators. To analyze the external propagator corresponding  to an   outgoing particle we should study the behavior of the function
 $$R_{a,\Omega}(k_,\sigma_1,k_2,\sigma_2,t)=\lim_{\tau\to\infty} \frac {\bra\theta|T(a(k_1,\sigma_1, \tau), a(k_2,\sigma_2,t)\hat S_{a,\Omega})|\theta\ket }{\bra\theta|\hat S_{a,\Omega}|\theta \ket }e^{i\epsilon(k_1)\tau}$$
 as $a\to 0.$ ( Under the sign of limit we have the two-point adiabatic Green function  constructed using operators $a(k,1)=a^+(k), a(k,-1)=a(k).$) This function can be represented in the form
 \begin{equation}\label{RR}\frac{\bra \theta|a(k_1,\sigma_1)\hat S_{a,\Omega}(\infty,t)e^{iH(0)_{\Omega}t} a(k_2,\sigma_2)e^{-itH(0)_{\Omega}t} \hat S_{a,\Omega}(t,-\infty)|\theta\ket)}{\bra \theta|\hat S_{a,\Omega}|\theta\ket}.\end{equation}
 It follows from adiabatic perturbation theory ( see the formula (\ref{PH})) that
 $$\lim e^{ic(t)}e^{-iH_{\Omega}(0)t}\hat S_{a,\Omega}(t,-\infty)\theta=\Theta_{\Omega}(g)$$
 
 $$\lim e^{ic(k,t)}\bra \theta|a(k_1,\sigma_1)\hat S_{a,\Omega}(\infty,t)e^{iH(0)_{\Omega}t}
 =\phi_{\Omega}(k,g)\delta_{-1}^{\sigma_1}$$
 where
 $$c(k,t)=\int_{-\infty}^0d\tau (E_{\Omega}(k,h(\tau)-\epsilon(k))+\int_0^td\tau E_{\Omega}(k,h(\tau),$$
 $$ c(t)=\int_{\infty}^ td\tau E_{\Omega}(h(\tau))$$
( We take the limit $a\to 0$ assuming that $h(t)=g.$)
 
 Using these formulas and (\ref{RR}) we obtain
 \begin{equation} \label {RRR}
 \lim e^{is_{\Omega}(k_1,h(t)} R_{a,\Omega}(k_,\sigma_1,k_2,\sigma_2,t)=\bra  \phi_{\Omega}(k_1,g)\delta_{-1}^{\sigma_1} |a(k_2,\sigma_2) |\Theta_{\Omega}(g)\ket.
 \end {equation}
 where $$s_{\Omega}(k,g)=\int_t^{\infty}d\tau\epsilon_{\Omega}(k,h(\tau)-\epsilon(k))+\epsilon(k)t.$$
 Again we take the limit $a\to 0$ assuming that $h(t)=g.$
 
 We considered external propagators for outgoing particles. External propagators for incoming particles can be calculated in the same way.
 
 Let us consider now Feynman diagrams for the right-hand side of (\ref{SSS}) assuming that $r_{a,\Omega}(k)$ is specified by the formula (\ref{R}). The vertices (1PI diagrams) and internal propagators are the same as for $\hat S_{a,\Omega}$  but the external propagators acquire phase factors.
 
 The formula (\ref {RRR})  gives the propagator corresponding to an external thick line in $p,t)$-representation for fixed volume $\Omega$ in these diagrams. One can check that the convergence to the limit in (\ref{RRR}) is uniform with respect to $\Omega$, therefore we can take the limit $\Omega\to\infty$ in this relation and obtain the external propagator in the infinite volume. To go to the $(p,\omega)$-representation we apply the stationary phase method to the Fourier transform. The stationary points are on the boundary $g=1.$  This  means that the limit $a\to 0$ is expressed in terms of 1PI diagrams on shell. It follows from (\ref{RR}) that in the limit $a\to 0, \Omega\to\infty$ matrix elements of
 $$ \frac{ \hat U_{a,\Omega}\hat S_{a,\Omega}\hat U_{a,\Omega}}{\bra\theta|\hat S_{a,\Omega}|\theta\ket}
$$ can be expressed in terms of amputated Green functions on shell; in other words,  we obtain LSZ formula for the scattering matrix. (We use the fact that the right-hand side of (\ref{RRR}) for $g=1$ coincides with the factor that appears in LSZ formula.)
 
 \section{Scattering matrix in the formalism of L-functionals}
 Let us consider formal translation-invariant  Hamiltonian (\ref{hh}) represented as a sum of quadratic Hamiltonian $\hat H(0)=\int \epsilon(k)a^+(k)a(k)dk$ and interaction Hamiltonian $\hat H_{int}=g\hat V$ that does not contain linear and quadratic terms. We assume that Feynman diagrams for adiabatic S-matrix $\hat S_{a, \Omega}$ do not have infrared and ultraviolet divergences after volume cutoff. 
 
 We will work with the corresponding "Hamiltonian"  $H=H(0)+gV$ in the formalism of L-functionals.  As we noticed in Section 2  in the situation we consider we have well-defined perturbation theory for adiabatic  S-matrix $S_a$  even without volume cutoff. More precisely, we can consider the space $\cal L$ of  functionals of the form 
  $$L(\alpha^*,\alpha)=\sum_k \sum_{m,n}  g^k\int dp^mdq^nL_{m,n,k}(\alpha^*(p_1)...\alpha^*(p_m)\alpha(q_1)...\alpha(q_n)$$
 where  the functions $L_{m,n,k}$ are smooth and fast decreasing at infinity. We assume that for every $k$ the sum over $m$ and $n$ is finite; in other words the coefficient functions of $L$ are polynomial in every order of perturbation theory.
 
 The space $\cal L$ contains, in particular, one-particle L-functionals 
 \begin{equation} \label {OP}
 L_f(\alpha^*,\alpha)=1-\int dpdp' f^*(p)f(p' )\alpha(p)\alpha^*(p').
 \end {equation}
 These functionals correspond to one-particle states in Fock space.
 
 The adiabatic matrix $S_a$ (as well as other operators of this kind)
 acts in the space  $\cal L$.  
 We define inclusive scattering matrix $S$ as an operator in space $\cal L$ that can be represented in the form
 \begin{equation}\label {IS}
 S=\lim U_aS_aU_a
 \end {equation}
 where
 $$U_a=e^{i \int dp s_a(p)(c^+_1(p)c_1(p)-c^+_2(p)c_2(p))}$$
 and  the function $s_a(p)$ is chosen in such a way that one-particle  L-functionals are $S$-invariant.  Namely, we will prove that one can take 
 \begin {equation}\label {UU}
 s_(p)=\int_{-\infty}^0d\tau(\epsilon (p,h(\tau))-\epsilon(p))
 \end {equation}
 where $\epsilon(p,g)$ are one-particle energies of the Hamiltonian $\hat H(g)=\hat H(0)+g\hat V$ (=poles of corresponding two-point Green function= poles of the two-point 
 GGreen function of the Hamiltonian"  $H(g).$
 
 It is easy to check  that the inclusive scattering matrix $S$ is related to the renormalized scattering matrix $\hat S$ by the formula
 \begin{equation}\label {HATS}
 (SL_K)(\alpha^*,\alpha)=L_{\hat SK\hat S^*}(\alpha^*,\alpha)
 \end{equation}
 hence $S$ contains the same amount of information as $\hat S$. The formula (\ref {HATS}) follows immediately from similar formula where $\hat S$ is replaced with $\hat S_{a,\Omega}.$  (see formula (\ref {HAT})).
 
 Inclusive cross-sections can be expressed  in terms of inclusive scattering matrix; this follows from (\ref{HATS}).
 
  We will sketch a proof of the existence of the operator $S$ satisfying these requirements (in the framework of perturbation theory).
 The proof  of the existence of the limit in (\ref{IS}) is based on the consideration of GGreen functions and adiabatic GGreen functions. These functions were defined at the end of Section 3; we deal with translation-invariant Hamiltonians therefore it is natural to include the dependence on spatial or momentum variables in these definitions.
 
 To define physical GGreen functions  in $(p,t)$-representation we  consider operators ${\bf c}(p,t,\sigma)=e^{Ht}c_{\sigma}(p)e^{-Ht}.$  We apply the chronological product 
 $T({\bf c}(p_1,t_1,\sigma_1)...{\bf c}(p_n,t_n,\sigma_n)$ to the L-functional describing translation-invariant state $\omega$; the GGreen function in state $\omega$ is defined as the value of  the functional $$T({\bf c}(p_1,t_1,\sigma_1)...{\bf c}(p_n,t_n,\sigma_n)\omega$$ at the point $\alpha^*=\alpha=0.$

 Adiabatic GGreen function is  defined as the value of the functional
 $$T(c(p_1,t_1,\sigma_1)... c(p_n,t_n,\sigma_n)e^{\int_{-\infty}^{\infty} d\tau h(a\tau) H_{int}(\tau)}) \omega_0$$
 at the point $\alpha^*=\alpha=0.$ (Here  $c(p,t,\sigma)=e^{H(0)t}c_{\sigma}(p)e^{-H(0)t}.$ 
 
 If the state $\omega$ is obtained from the state $\omega_0$ by adiabatic dressing then the adiabatic GGreen function tends to physical GGreen function as  $a$ tends to zero.
 
 We are interested in physical GGreen functions for the case when $\omega$ is the ground state; then we should take $\omega_0=1.$
 
 As we noticed one can construct diagram techniques for the calculation of the adiabatic  S-matrix $S_a$, adiabatic GGreen functions, and physical GGreen functions.
 Propagators in these techniques are two-point GGreen functions of the ''Hamiltonian" $H(0)$. However,  as in the preceding section, we will use a modification of these techniques where propagators are physical two-point GGreen functions and vertices are 1PI diagrams.
 It is convenient to work in $(p,\omega)$-representation. Then again we can assume that vertices (=1PI diagrams) and internal propagators have no poles and taking the limit $a\to 0$  in their contributions to the adiabatic S-matrix  $S_a$  we get the contribution of vertices and internal propagators to GGreen functions. The situation with external propagators in the diagrams for $S_a$ is more complicated. To get a finite limit when $a\to 0$ we should consider diagrams
 for  $U_aS_aU_a$ instead of diagrams for $S_a$; this does not change vertices and internal propagators but changes external propagators. In the definition of $U_a$ we should take
 $s_a(p)$ given by the formula (\ref{UU}). Then we can calculate  external propagators
 using the methods of the preceding section. The same considerations show that the limit  of diagrams for $U_aS_aU_a$ as $a\to 0$ can be  expressed in terms of  1PI diagrams on shell. This means that in the framework of perturbation theory  we can say that matrix elements of $U_aS_aU_a$ in the  limit  $a\to 0$ can be expressed in terms of amputated GGreen functions on shell.

 \section {$\hbar\to 0$}
 
 We noticed already that equations of motion for L-functionals and adiabatic scattering matrix in the formalism of L-functionals have a limit  as $\hbar\to 0.$ It follows that the same statement is correct  for inclusive scattering matrix defined as scattering matrix in the formalism of L-functionals.
 
 Let us show that inclusive scattering matrix in  $\hbar \to 0$ limit can be described in terms of GGreen functions 
 in Keldysh (physical) basis. For simplicity we will work with Hermitian scalar field $\phi (\bx,t)$.  Two operators induced by this field in the space $\cal L$ of functionals $L(\alpha^*,\alpha)$ will be denoted by 
 $\phi_l (\bx,t)$ and $\phi_r(\bx,t)$. To define GGreen functions in the ground state $\omega$ we apply  a chronological product of operators
  $\phi_l (\bx_i,t_i)$ and $\phi_r(\bx_j,t_j)$ to the functional $\omega (\alpha^*, \alpha)$ and calculate the value of the functional we obtained at the point $\alpha^*=\alpha=0$.  We can change the basis in the space of 
  GGreen functions  introducing operators $\phi_{qu}(\bx, t)=
  \phi_l(\bx, t)-
  \phi_r(\bx,t)$, $\phi_{cl}(\bx,t)=\frac {1}{2} (\phi_l(\bx, t)+
  \phi_r(\bx,t))$ (Keldysh basis). Operators $\phi(\bx, t)$ are Hermitian linear combinations of creation and annihilation operators; this means that operators  $\phi_{qu}(\bx, t)$ are proportional to $\hbar$ as follows from 
  (\ref{BBB}). (This justifies the subscript $qu$ that stands for quantum.
  The subscript $cl$ stands for classical.) GGreen functions in Keldysh basis
  are labelled by indices $qu$ and $cl$. One can check that the limit of inclusive scattering matrix as $\hbar\to 0$ is governed by GGreen function with only one index $qu$.
  
{\bf Acknowledgements. } I appreciate the hospitality of IHES,IAS, Simons Center, Caltech and M.Kontsevich, E. Witten, L. Alvarez-Gaume, N. Nekrasov, A.Kapustin.
This work was supported by NSF grant PHY-2207584.

\end {document}